\documentclass[journal]{IEEEtran}
\IEEEoverridecommandlockouts
\usepackage{cite}
\usepackage{amsmath,amssymb,amsfonts}
\usepackage{algorithmic}
\usepackage{graphicx}
\usepackage{textcomp}
\usepackage{xcolor}
\usepackage{cite}
\usepackage{float}
\usepackage{glossaries}
\def\BibTeX{{\rm B\kern-.05em{\sc i\kern-.025em b}\kern-.08em
		T\kern-.1667em\lower.7ex\hbox{E}\kern-.125emX}}

\newacronym{SFTR}{SFTR}{Single Fluctuating Two-ray}
\newacronym{IFTR}{IFTR}{independent fluctuating two-ray}
\newacronym{FTR}{FTR}{fluctuating two-ray}
\newacronym{PDF}{PDF}{probability density function}
\newacronym{SNR}{SNR}{Signal to Noise Ratio}   
\newacronym{CDF}{CDF}{cumulative distribution function}
\newacronym{GMGF}{GMGF}{generalized moment generating function}

\newtheorem{corollary}{Corollary}

\newtheorem{lemma}{Lemma}
\newtheorem{definition}{Definition}

\newtheorem{remark}{Remark}

\begin{document}
	\bstctlcite{IEEEexample:BSTcontrol}
	
	\title{
		A Tractable Statistical Representation of IFTR Fading with Applications
		}
		
	\author{
		Maryam Olyaee, Hadi Hashemi and Juan~M.~Romero-Jerez, \emph{ Senior Member, IEEE}
		\thanks{
		This work was submitted to the IEEE for publication. Copyright may be transferred without notice, after which this version may no longer be accessible.}
		\thanks{
This work has been
funded in part by Junta de
Andalucía through project P21-00420 and also grant EMERGIA20-00297, and in part by
MCIN/AEI/10.13039/501100011033 through grant PID2020-118139RB-I00.}
\thanks{
M. Olyaee and J. M. Romero-Jerez are with Communications and Signal Processing Lab, Telecommunication Research Institute (TELMA), Universidad de M\'alaga, 
ETSI Telecomunicaci\'on, Bulevar Louis Pasteur 35, 29010 M\'alaga, Spain. 
Hadi Hashemi is with the Department of Signal Theory, Networking and Communications, Universidad de Granada, 18071, Granada, Spain.
(e-mails: maryam.olyaee@ic.uma.es, romero@dte.uma.es, hashemi@ugr.es,).}
}

	\maketitle
	
	\begin{abstract}
The recently introduced \gls{IFTR} fading model, consisting of two specular components fluctuating independently plus a diffuse component, has proven to provide an excellent fit to different wireless environments, including the millimeter-wave band. However, the original formulations of the probability density function (PDF) and cumulative distribution function (CDF) of this model are not applicable to all possible values of its defining parameters, and are given in terms of multifold generalized hypergeometric functions, which prevents their widespread  use for the derivation of performance metric expressions. In this paper we present a new formulation of the IFTR model as a countable mixture of Gamma distributions which greatly facilitates the performance evaluation for this model in terms of the metrics already known for the much simpler and widely used Nakagami-$m$ fading. Additionally, a closed-form expression is presented for the generalized moment generating function (GMGF), which permits to readily obtain all the moments of the distribution of the model, as well as several relevant performance metrics. Based on these new derivations, the \gls{IFTR} model is evaluated for the average channel capacity, the outage probability with and without  co-channel interference, and the bit error rate (BER), which are verified by Monte Carlo simulations.
	\end{abstract}
	
\glsreset{IFTR}
\glsreset{PDF}
\glsreset{CDF}
\glsreset{SNR}
\glsreset{GMGF}
	
	\section{Introduction}
	Due to the spectrum shortage in future generation wireless network, higher frequency bands are being considered in the standards in order to cover users' demands. Thus, millimeter and terahertz bands usage has emerged in the context of 5G/6G cellular networks \cite{6736746}. In many wireless scenarios, channel multipath fading is an essential propagation effect to be considered due to the potential detrimental impact on performance. Therefore, accurate characterization of wireless channel fading at those higher frequencies has become a relevant research topic, and much effort is being made in this area \cite{marins2021fading,uwaechia2020,6515173}.
	
	Recently, the \gls{IFTR} \cite{IFTR} channel model has been presented to characterize multipath propagation, which includes several well-known distributions, namely  Rayleigh, Rician, Hoyt (Nakagami-$q$), Rician Shadowed, and Nakagami-$m$,  as special or limiting cases. The IFTR model consists of two dominant (specular) waves plus a diffuse component, due to the aggregation of multiple low-power scattered waves, modeled as a complex Gaussian random variable (RV), where the specular components are assumed to fluctuate independently following Nakagami-$m$ fading. This model is related to the \gls{FTR} fading model except that in the latter the two specular components are assumed to be fully correlated and fluctuate simultaneously. The FTR model was introduced in \cite{romero2017fluctuating} and was later reformulated in \cite{zhang2017new,lopez2021comments} and, more recently, in \cite{NewFTR}, and has been studied abundantly for different wireless environments, mostly in the context of millimeter-wave communications, and considering many different performance metrics (see for example \cite{NewFTR} and the references therein). In spite of the apparent similitude in the formal definition of the  FTR and IFTR fading models, there are major differences between them, both in terms of the fitting results to experimental measurements and in the involved mathematical derivations. On the one hand, the IFTR fading model has been shown to provide a (sometimes remarkable) better fit than FTR fading (as well as other generalized fading models such as $\kappa$-$\mu$ shadowed \cite{Paris14} and two-wave with diffuse power --TWDP-- \cite{NewFTR}) to experimental data in very different environments, including line-of-sight (LOS) millimeter-wave, land-mobile satellites (LMS), and underwater acoustic communications (UAC) \cite{IFTR}. On the other hand, the independence of the two specular components in the IFTR model imposes new mathematical challenges, as now a two-fold nested integration always appear in its statistical characterization.
	
	Although both the \gls{PDF} and \gls{CDF} of the IFTR model were presented in \cite{IFTR}, their use is rather limited for two reasons: on the one hand they are not completely general, as they require assuming one of the model parameters $m_1$ or $m_2$ to be integer, while they can take any arbitrary positive real value in realistic propagation scenarios; on the other hand, the known PDF and CDF are given in terms of a generalized hypergeometric function, which is actually a multifold infinite summation, which is very difficult to manipulate to obtain analytical expressions for most performance metrics in wireless communication systems.

	In this paper, we solve the aforementioned issues by deriving a new statistical characterization of the \gls{IFTR} fading model assuming arbitrary positive values of $m_1,m_2$ and easy to manipulate. Additionally, we expand the known results for the precise characterization of the model and apply them for the performance analysis of wireless systems.  
	Specifically, the key contributions of this paper are:
	\begin{itemize}
		\item A new formulation is presented for the PDF and CDF of the instantaneous SNR of IFTR fading in terms of an infinite countable mixture of Gamma distributions for arbitrary values of the channel parameters $m_1$ and $m_2$, where the weights of the elements of the mixture are given in closed-form. The resulting infinite series are demonstrated to be convergent and are precisely truncated and evaluated using the Kolmogorov-Smirnov goodness-of-fit test.
		\item The \gls{GMGF} of the IFTR fading model is obtained for the first time, which for many relevant cases can be written in closed-form, allowing to obtain all the moments of the distribution. In spite of the model generality and statistical complexity, this function permits to obtain closed-form expressions for different relevant performance metrics including, for example, secrecy capacity outage, outage probability under  interference and energy detection probability.
		\item
		The new and expanded statistical characterization of IFTR fading is used for its performance analysis evaluation in terms of the average capacity, outage probability with and without interference and average bit error rate (BER) for different modulations. The effect of the parameters values of the model are evaluated numerically and verified by simulation.
	\end{itemize}

	The rest of this paper is organized as follows:
The channel model is presented in Section II.
Then, in Section III, the new representation of the \gls{IFTR} fading is presented, as well as, for the first time, to the authors' knowledge, an expression of the \gls{GMGF}, which for many relevant cases can be written in closed-form. 
Several performance metrics, including the average capacity, the outage probability, and the BER in \gls{IFTR} fading are analyzed in Section IV. Simulation and numerical results are given in Section V. Finally, the paper is concluded in Section VI.

	\section{Preliminary definitions and channel model}
	
	\begin{definition}
 A RV $X$ following a  Gamma distribution with shape parameter $\lambda$ and scale parameter $\nu$ will be denoted as $X \sim \mathcal{G}(\lambda,\nu)$, and its PDF and CDF will be given, respectively, by
\begin{align} \label{fg}
	&f^{\mathcal{G}}(x;\lambda,\nu)=\frac{x^{\lambda-1}}{\Gamma(\lambda)\nu^{\lambda}}e^{-\frac{x}{\nu}},\\&
\label{FFg}
	F^{\mathcal{G}}(x;\lambda,\nu)=\frac{1}{\Gamma(\lambda)} \gamma\left(\lambda,\frac{x}{\nu}\right),
\end{align}
where $\gamma(\cdot,\cdot)$ is the incomplete Gamma function \cite[eq. (8.350.1)]{gradshteyn2014table}.
	\end{definition}
	
	\begin{remark}
The SNR $\gamma_\mathcal{K}$ (or, equivalently, the received power) in a Nakagami-$m$ fading with mean $\bar\gamma_\mathcal{K}$ and fading severity parameter $m$ follows a  Gamma distribution with shape parameter $m$ and scale parameter $\bar\gamma_\mathcal{K}/m$, i.e., 
$\gamma_\mathcal{K} \sim \mathcal{G}(m,\bar\gamma_\mathcal{K}/m)$.
	\end{remark}

The IFTR fading model is composed of two specular waves, whose amplitude fluctuate according to independent	Nakagami-$m$ fading, plus an undetermined number of scattered low-amplitude waves (the diffuse component) which, by virtue of the central limit theorem, are jointly represented by a complex Gaussian RV. Let $\zeta_i \sim \mathcal{G}(m_i,1 /m_i)$, with $i \in \{1,2\}$, then the complex base-band representation of the IFTR fading model can be expressed as
\begin{align}\label{eq1}
	V_r = \sqrt{\zeta_1} V_1 e^{j \phi_1} + \sqrt{\zeta_2} V_2 e^{j \phi_2} + X + j Y, 
\end{align}
where $V_i$ is the average amplitude of the $i$-th specular component, $\phi_i$ is a  uniformly distributed RV in $[0,2\pi)$ representing its phase,
and $X + j Y$ models the diffuse component with $X,Y \sim \mathcal{N}(0,\sigma^2)$. 

In addition to the fading severity parameters of the specular components, $m_1$ and $m_2$, the IFTR model will be determined by the following physically-motivated parameters:
\begin{align}
	K = \frac{V_1^2+V_2^2}{2\sigma^2},\\
	\Delta = \frac{2V_1V_2}{V_1^2+V_2^2},
\end{align}
where $K$ represents the ratio of the average power of the dominant components to the power of the diffuse component and $\Delta \in [0,1]$ provides a measure of the specular components similarity, so that $\Delta=0$ implies $V_1=V_2$. Without loss of generality we will assume $V_1 \geq V_2$, and therefore $\Delta=1$ implies $V_2=0$, i.e., only the first specular component, if any, is received. For the sake of compactness in subsequent expressions, we will also define the following ancillary parameters, given in terms of $K$ and $\Delta$: 
	\begin{align}
	{K_1} \triangleq \frac{{V_1^2}}{{2{\sigma ^2}}} = K\frac{{1 + \sqrt {1 - {\Delta ^2}} }}{2},
	\\
	{K_2} \triangleq \frac{{V_2^2}}{{2{\sigma ^2}}} = K\frac{{1 - \sqrt {1 - {\Delta ^2}} }}{2}.
\end{align}

The IFTR model is very versatile and includes different classical and generalized fading models as particular cases by an appropriate selection of the parameters. Thus, for $m_1,m_2 \rightarrow \infty$ the fluctuations of the specular components tend to disappear and the IFTR model collapses to the TWDP one \cite{Durgin02}. If, in addition, we let $\Delta=0$, the Rice model is obtained. For finite values of $m_1$, $\Delta=0$ yields the Rician Shadowed model \cite{abdi2003new}, which was shown in \cite{moreno2016kappa} that it contains the Hoyt (Nakagami-$q$) model for $m_1=0.5$, with $q=\left(\sqrt{1+2K}\right)^{-1}$. The Rayleigh fading model can be obtained as a particularization of either the aforementioned Rice or Hoyt models for $K=0$, and also for $m_1=1$ and  $\Delta=0$. If there is only one specular component and the diffuse component is absent ($\Delta=0$, $K\rightarrow \infty$), the IFTR model collapses to the Nakagami-$m$ model.

\section{New representation of the \gls{IFTR} fading model}
In this paper, we present a new statistical characterization of the SNR of a signal undergoing IFTR fading which, denoting by $E_s$ the symbol  energy density and $N_0$ the power spectral density, is defined as $\gamma \triangleq {(E_s/N_0)\left|V_r\right|^2}$. 

\begin{definition}
A RV $\gamma$ following an IFTR distribution with parameters $m_1$, $m_2$, $K$, $\Delta$ and mean $\overline\gamma$ will be denoted by 
$\gamma \sim \mathcal{IFTR}(\overline{\gamma},m_1, m_2,K,\Delta)$, and its PDF and CDF will be denoted, respectively, by $f_{\gamma}^{\rm IFTR}(\cdot)$ and $F_{\gamma}^{\rm IFTR}(\cdot)$.
\end{definition}

Following the same spirit as in \cite{ermolova2016capacity} for TWDP and in \cite{zhang2017new,lopez2021comments} for FTR fading, we now show that the PDF and CDF of the SNR of a RV following an IFTR distribution can be expressed as infinite countable mixtures of the corresponding functions for the Gamma distribution. Additionally, we show how this result can be applied to readily obtain any metric, defined by averaging over the channel realizations, for the IFTR model, from such metric for the much simpler and widely used Nakagami-$m$ fading.

\subsection{\gls{PDF} and \gls{CDF} of IFTR fading}
\begin{lemma}
 Let $\gamma \sim \mathcal{IFTR}(\overline{\gamma},m_1, m_2,K,\Delta)$, then, its PDF and CDF can be expressed, respectively, as 
	\begin{align}\label{pdfI}
		&	{f_{\gamma}^{\rm IFTR}}\left( x \right) = \sum_{j = 0}^{\infty}  A_j {f^{\mathcal{G}}}\left( {x;j + 1,\frac{\bar\gamma}{1+K}} \right),
\\  &	{F_{\gamma}^{\rm IFTR}}\left( x \right) = \sum_{j = 0}^{\infty}  A_j {F^{\mathcal{G}}}\left( {x;j + 1,\frac{\bar\gamma}{1+K}} \right),\label{cdfI}
	\end{align}
	where  $f^{\mathcal{G}}$ and $F^{\mathcal{G}}$ are, respectively, the PDF and CDF of the Gamma distribution given in \eqref{fg} and \eqref{FFg}, and coefficients $A_j$ are given in \eqref{eq3} in terms of the channel parameters and the regularized Gauss hypergeometric function\footnote{The regularized Gauss hypergeometric function can be calculated in terms of the \textit{standard} Gauss hypergeometric function as 
	$_2 \tilde F_1 \left( {a,b;c;z} \right) = \,_2 F_1 \left( {a,b;c;z} \right)/\Gamma (c)$ when $c \notin \{0,-1,-2,\ldots \}$, however, the corresponding parameter $c$ in the coefficients $A_j$ in (\ref{eq3}) can indeed be a non-positive integer for some values of index $j$, therefore, $_2 \tilde F_1$ has to be calculated using (\ref{regularized 2F1}). Nevertheless, the regularized Gauss hypergeometric function is in-built in the Mathematica software.}, which is defined as
		\begin{align} \label{regularized 2F1}
_2 \tilde F_1 \left( {a,b;c;z} \right) = \sum\limits_{k = 0}^\infty  {\frac{{\left( a \right)_k \left( b \right)_k }}
{{\Gamma \left( {c + k} \right)}}} \frac{{z^k }}{{k!}},
	\end{align}
	where $(a)_k\triangleq \Gamma (a+k) / \Gamma (a)$ is the Pochhammer symbol.
\end{lemma}
\begin{IEEEproof}
	See Appendix A.
\end{IEEEproof}	
	
	Note that, in contrast to the PDF and CDF expressions given in \cite{IFTR}, \eqref{pdfI} and \eqref{cdfI} are valid for arbitrary values of $m_1$ and $m_2$, and therefore this is also true for all the performance metrics derived from them.
	
	\begin{figure*}[t]
\begin{align}\label{eq3}
    A_j =& \sum_{k=0}^{j}\binom{j}{k} \sum_{q=0}^{j-k} \binom{j-k}{q}  \frac{K_1^q K_2^{j-k-q} }{j!} \sum_{l=0}^{k} \binom{k}{l} \frac{m_1^{m_1}}{\Gamma(m_1)} \frac{m_2^{m_2}}{\Gamma(m_2)} \frac{\Gamma(m_1+q+l)}{(K_1+m_1)^{m_1+q+l}} \frac{\Gamma(m_2+j-{k}-q+l)}{(K_2+m_2)^{m_2+j-{k}-q+l}} \nonumber\\
    &\times (-1)^{k}{\left(\frac{K 	\Delta}{2}\right)^{2l}}{} 	{}_2\tilde{F}_1\left( m_1+q+l ,m_2+j-{k}-q+l,2l-k+1,\frac{K^2\Delta^2}{4(K_1+m_1)(K_2+m_2)}\right). 
    \\ \nonumber
    \\ \hline \nonumber 
\end{align}
\end{figure*}

\begin{remark}
By noting that the $j$-th term in \eqref{pdfI} is proportional to $(x/\bar\gamma)^{j}$, the PDF and CDF in IFTR fading in the high SNR regime (i.e., as $\bar\gamma\rightarrow\infty$) can be approximated by only maintaining the first term in the infinite summations, yielding
\begin{align}\label{pdfIapp}
  &	{f_{\gamma}^{\rm IFTR}}\left( x \right) \approx A_0 \frac{{\bar \gamma }}
{{1 + K}}e^{ - x(1 + K)/\bar \gamma } , \ \, \quad\quad\quad \bar \gamma  \gg x , \hfill \\
 &	{F_{\gamma}^{\rm IFTR}}\left( x \right) \approx A_0 \left( {1 - e^{ - x(1 + K)/\bar \gamma } } \right),\quad\quad\quad \hfill \bar \gamma  \gg x
\end{align}
with
\begin{align}\label{A0}
  A_0  =& \frac{{m_1^{m_1 } m_2^{m_2 } }}
{{\left( {K_1  + m_1 } \right)^{m_1 } \left( {K_2  + m_2 } \right)^{m_2 } }}\; \hfill \nonumber\\&
   \times \;_2 F_1 \left( {m_1 ,m_2 ;1;\frac{{K^2 \Delta ^2 }}
{{4\left( {K_1  + m_1 } \right)\left( {K_2  + m_2 } \right)}}} \right). 
\end{align}
\end{remark}

\begin{corollary} \label{C1}
	Let $h(\gamma)$ be a performance metric (or statistical function) depending on the instantaneous SNR, and let $X_\mathcal{K}
	(\bar\gamma_\mathcal{K},m)$ be the metric (or function) obtained 	by averaging over an interval of the PDF of the SNR for Nakagami-$m$ fading with mean $\bar\gamma_\mathcal{K}$ and fading severity $m$, i.e.,	
	\begin{align} \label{XK}
		X^\mathcal{K}(\bar\gamma_\mathcal{K},m) = \int_{a}^{b} h(x)f^\mathcal{G}(x;m,\bar\gamma_\mathcal{K}/m)dx,
	\end{align}
	where $0 \leq a \leq b < \infty$. Then, the average performance metric for \gls{IFTR} fading can be calculated as
	\begin{align} \label{XIF}
		X^{\rm IFTR}&\left(\bar\gamma, m_1,m_2,K,\Delta\right) \nonumber\\&= \sum_{j=0}^{\infty} A_j X^\mathcal{K}\left(\frac{\bar\gamma}{1+K}(j+1),j+1\right),
	\end{align}
	where $A_j$ are the \gls{IFTR} coefficients defined in \eqref{eq3}.
\end{corollary}
\begin{IEEEproof}
	The average metric in \gls{IFTR} fading channel is calculated as
	\begin{align}\label{XS}
		X^{\rm IFTR}(\bar\gamma,m_1,m_2,K,\Delta) = \int_{a}^{b} h(x)f_{\gamma}^{\rm IFTR}\left( x \right) dx .
	\end{align}
	By plugging \eqref{pdfI} into \eqref{XS} we can write
	\begin{align}\label{XS2}
  &X^{\rm IFTR}\left( {\bar \gamma  ,m_1 ,m_2 ,K,\Delta } \right) \hfill \nonumber\\&
  \quad  = \int_a^b {h\left( x \right)} \left[ {\sum\limits_{j = 0}^\infty  {A_j f^\mathcal{G} \left( {x;j + 1,\frac{{\bar \gamma  }}
{{1 + K}}} \right)} } \right]dx \hfill \nonumber\\&
  \quad  = \sum\limits_{j = 0}^\infty  {A_j } \int_a^b {h\left( x \right)} f^\mathcal{G} \left( {x;j + 1,\frac{{\bar \gamma  }}
{{1 + K}}} \right)dx .
\end{align}
Comparing the integral of the resulting expression with \eqref{XK} and identifying $j + 1=m$ and $\frac{{\bar \gamma  }}
{{1 + K}} = \frac{\bar \gamma_\mathcal{K}  } {m}$, \eqref{XIF} is obtained.
\end{IEEEproof}

\subsection{Series convergence and Kolmogorov-Smirnov goodness-of-fit statistical test}
\label{KS}
The series expressions of the PDF given in \eqref{pdfI} is calculated by averaging the convergent series expression for TWDP fading, given in \eqref{eq:20}, over the fluctuations of the specular components, as explained in Appendix A. The weights of the Gamma PDF's in the TWDP series are positive \cite{ermolova2016capacity} and therefore the interchange of integration and infinite summation in \eqref{eq:26} can be carried out by virtue of Tonelli’s theorem \cite{vetterli2014foundations}, which has the following consequences:

(i) The series in the right hand side of \eqref{pdfI} converges to  the PDF of the IFTR fading model ${f_{\gamma}^{\rm IFTR}}(x)$ $\forall  x \in [0,\infty)$.

(ii) The calculated coefficients $A_j$ are positive for all $j$.

Moreover, the performance metrics in communication systems (e.g., BER, channel capacity, outage probability, etc.) are typically non-negative functions which, together with (ii), permits to invoke again Tonelli’s theorem, thus allowing the interchange of integration and infinite summation in \eqref{XS2}, yielding two additional consequences:

(iii) The series in the right hand side of \eqref{XIF} converges to the average metric in IFTR fading $X^{\rm IFTR}(\bar\gamma,m_1,m_2,K,\Delta)$.

(iv) Considering $h(\gamma)=1$ in $[0,\infty)$ in Corollary 1 yields $\sum\limits_{j=0}^\infty  {A_j  = 1}$. Adittionally, considering $h(\gamma)=1$ in $[0,x)$ in Corollary 1 provides a formal justification for obtaining \eqref{cdfI} by integrating \eqref{pdfI} term by term.

The infinite series used in the statistical characterization of IFTR fading must be truncated for numerical computation.
We now provide the Kolmogorov-Smirnov (KS) goodness-of-fit statistical test, which permits to check how close a truncated series is to the exact value. The KS test statistic is given by \cite{papoulis1994random}
\setcounter{equation}{37}
\begin{align}
	T_{KS} = max |\hat{F}_{\gamma}^{\rm IFTR}(x)-F_{\gamma}^{\rm IFTR}(x)|,
\end{align}
where $F_{\gamma}^{\rm IFTR}(x)$ is the exact value of the \gls{CDF} and $\hat{F}_{\gamma}^{\rm IFTR}(x)$ is the approximation of the \gls{CDF} when the series is truncated to $J$ terms.
\begin{table}
	\caption{KS test for \gls{IFTR} channel with different channel parameters $K$, $m_1$, $m_2$, $\Delta$} \label{T2}
	\centering
	\begin{tabular}{c|c|c|c}
		\hline
		\hline
		channel parameters & $J=20$ & $J=30$ & $J=40$ \\
		\hline
		\hline
		$K=10$, $m_1=8$, $m_2=5$, $\Delta=0.5$ & 0.0516 & 0.0017 & 0.0013 \\
		\hline
		$K=15$, $m_1=8$, $m_2=5$, $\Delta=0.5$ & 0.2331 & 0.0465 & 0.0052 \\
		\hline
		$K=10$, $m_1=8$, $m_2=5$, $\Delta=0.8$ & 0.0831 & 0.0055 & 0.0043 \\
		\hline
		$K=10$, $m_1=8$, $m_2=5$, $\Delta=0.3$ & 0.0364 & 0.0008 & 0.0007 \\
		\hline
		$K=10$, $m_1=15$, $m_2=5$, $\Delta=0.5$ & 0.0388 & 0.0011 & 0.0009 \\
		\hline
	\end{tabular}
\end{table} 

Table \ref{T2} reports the KS test for different channel parameters when the truncated series have 20, 30, or 40 terms. 
It can be seen that the accuracy reaches an acceptable level when the first 40 terms of the series are computed, so the numerical calculations of all the series in this work will consider 40 terms.

\begin{figure}[t]
    \centering
    \includegraphics[height=7cm]{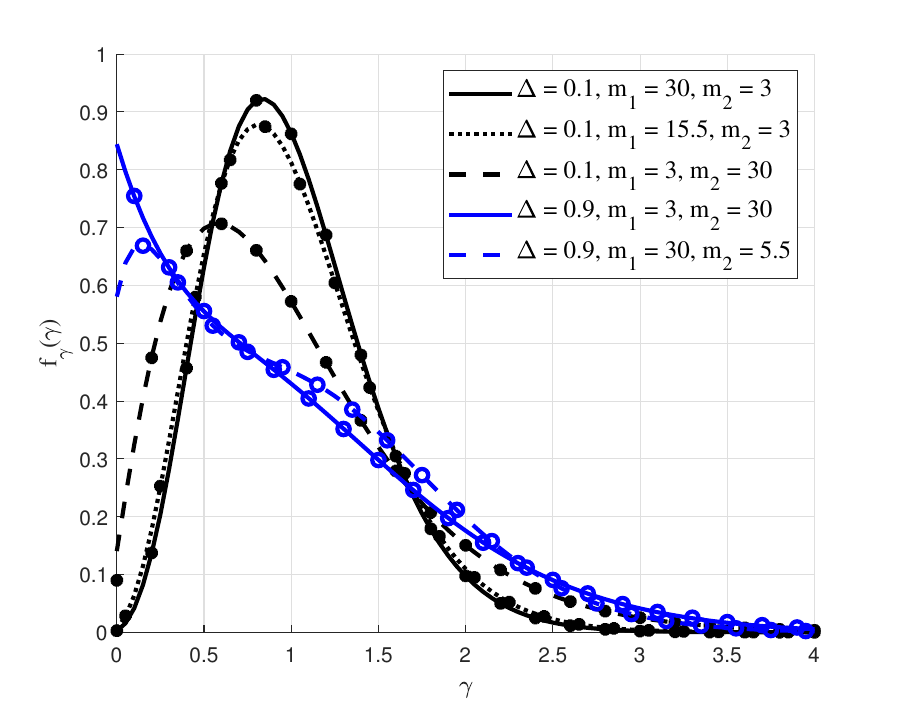}
    \caption{PDF of the SNR under IFTR fading  for different channel parameters $m_1,m_2$ and $\Delta$. Simulation confirmation results are displayed as circular markers. $K=10$. $\bar\gamma=1$.}
    \label{figure1}
\end{figure}

\begin{figure}[t]  
\centering\includegraphics[height=7cm]{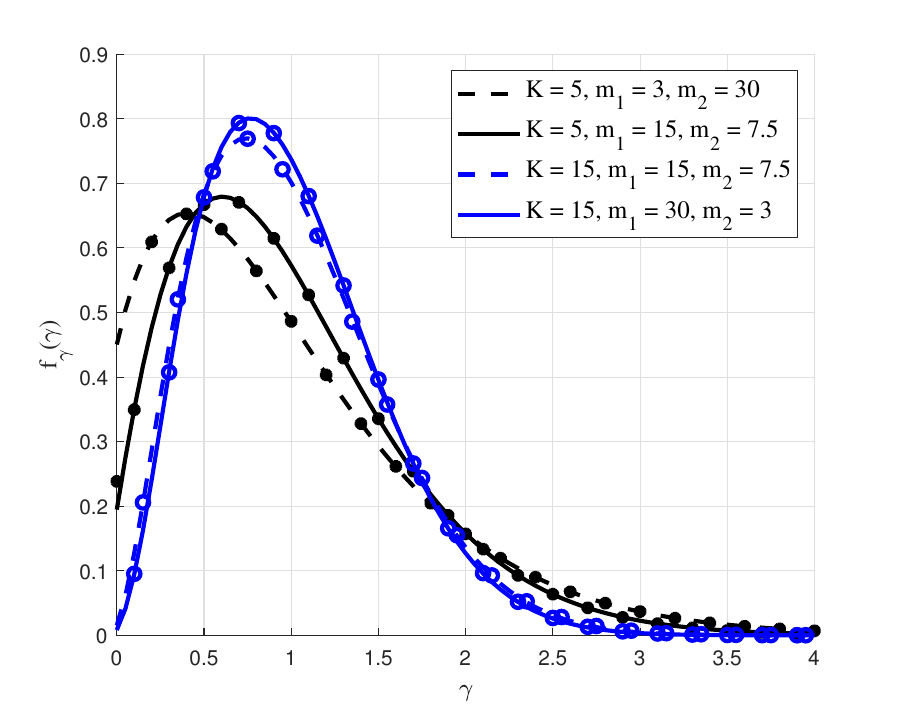}
    \caption{PDF of the SNR under IFTR fading for different channel parameters $m_1,m_2$ and $K$. Simulation confirmation results are displayed as circular markers. $\Delta=0.5$. $\bar\gamma=1$.}
    \label{figure2}
\end{figure}

\begin{figure}[t]
	\centering
	\includegraphics[height=7cm]{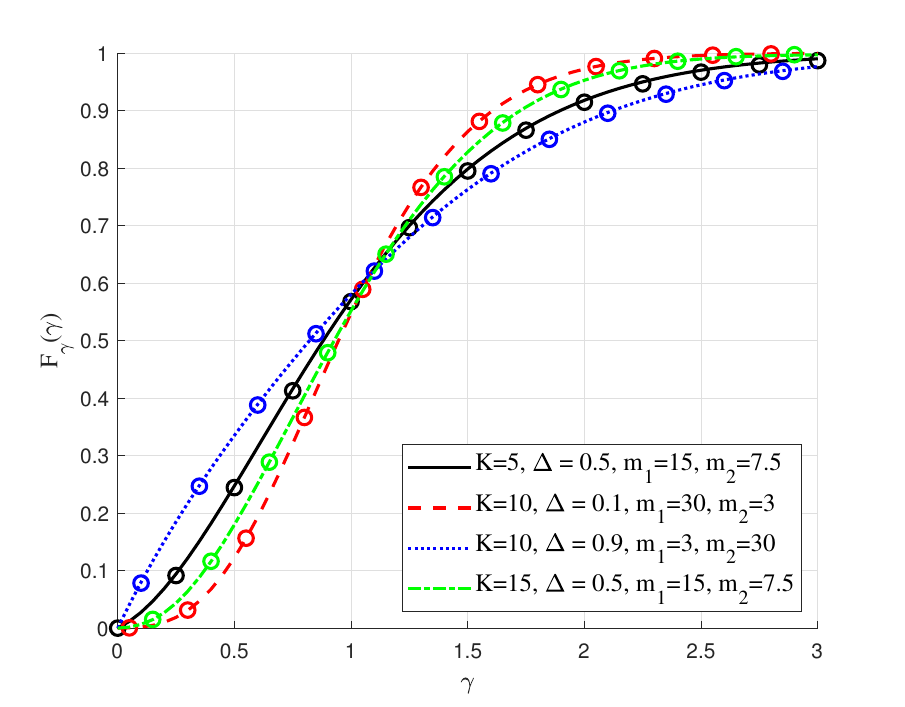}
	\caption{CDF of the SNR under IFTR fading  for different channel parameters $m_1,m_2$, $K$, and $\Delta$. Simulation confirmation results are displayed as circular markers.}
		\label{figure3}
\end{figure}

Figs. \ref{figure1} and \ref{figure2} show the PDF of the SNR for different IFTR channel parameters obtained from \eqref{pdfI} assuming 40 terms in the truncated series computation. Fig. \ref{figure1} is plotted for $\Delta=0.1,0.9$ and for both integer and non-integer values of $m_1$ and $m_2$, while Fig. \ref{figure2} shows the PDF for $K=5,15$. The numerical results are verified by Monte-Carlo simulation, showing an excellent agreement in all cases. 
Fig. \ref{figure3} illustrates the CDF  of the SNR in IFTR fading computed from \eqref{cdfI} for different values of $K$, $\Delta$ and $m_1,m_2$. 

\subsection{GMGF and moments of the IFTR model}
\begin{definition}
Let $n>0$, and let $X$ be a continuous non-negative RV with PDF $f_X(\cdot)$. The GMGF of $X$ is defined as
\begin{equation} \label{defgmgf}
\phi _X ^{(n)} \left( s \right) \triangleq  E\left\{ {X ^n e^{X s} } \right\} = \int_0^\infty  {x^n e^{xs} f_X  } \left( x \right)dx,
\end{equation}
where $E\left\{\cdot\right\}$ denotes the expectation operator.
The moment generating function (MGF) is defined as $\phi _X  \left( s \right) \triangleq E\left\{ {e^{X s} } \right\} = \phi _X^{(0)} \left( s \right)$, and it is therefore a particular case of the GMGF. Note that for $n\in \mathbb{N}$, the GMGF coincides with the $n$-th order derivative of the MGF. Also, the $n$-th order moment of $X $ is  obtained as $\mu _X^n  \triangleq E\left\{ {X^n } \right\} = \phi _X ^{(n)} \left( 0 \right)$.
\end{definition}

The GMGF finds application in different communication theory areas, including energy detection, outage probability
under co-channel interference, physical layer security or BER analysis. In most cases it suffices to consider $n\in \mathbb{N}$, which usually results in closed-form expressions for the GMGF, such as it is the case for IFTR fading, as we show bellow. However, there are situations, such as composite Inverse Gamma (IG) shadowing/fading modeling \cite{ramirez2021composite}, where the more general case of arbitrary $n>0$ needs to be considered.
In the following Lemma we derive expressions for the GMGF of the IFTR fading model for both cases.

\begin{lemma}
	Let $\gamma \sim \mathcal{IFTR}(\overline{\gamma},m_1, m_2,K,\Delta)$, then, its GMGF can be expressed as follows:
	
	(i) General case ($n\in \mathbb{R^+}$):
	\begin{align}
			\phi_{\gamma}^{(n)} (s) 
		= \sum_{j=0}^{\infty} A_j \phi_{G}^{(n)} \left(s,j+1,\frac{\bar\gamma}{1+K}\right),\label{gmgfs}
	\end{align}
where $A_j $ is defined in \eqref{eq3} and $\phi_{G}^{(n)}$ is the \gls{GMGF} of a RV $G \sim \mathcal{G}(\lambda,\nu)$, which is given by
    \begin{align} \label{GMGFG}
	\phi_{G}^{(n)} (s,\lambda,\nu) = \frac{\Gamma(n+\lambda) \left( \frac{ 1 }{\nu }-s\right)^{-(n+\lambda)}}{\Gamma(\lambda) \nu^{\lambda}}  .
\end{align}

    (ii) Case $n\in \mathbb{N}$: A closed-form expression is given in \eqref{iftr_gmgf}.
\end{lemma}			

		\begin{IEEEproof}
		
		Case (i): This result is obtained by 
		by applying Corollary \ref{C1} to the GMGF of the SNR in Nakagami-$m$ fading given in  \cite[Table II]{ramirez2021composite}.
		
		Case (ii): See Appendix B.
	
	\end{IEEEproof}

\begin{figure*}
    \begin{align}\nonumber
    &\phi _{\gamma} ^{(n)} \left( s \right)=	\frac{\overline \gamma  ^n n!}{\left(1+K-\bar{\gamma }s\right)^{n+1-{m_1}-{m_2}}} \frac{{{m_1^{m_1}}}}{{\Gamma (m_1)}} \frac{{{m_2^{m_2}}}}{{\Gamma (m_2)}} \sum_{q = 0}^n \binom{n}{q} \frac{{{\left( 1 + K \right)^{q+1} }}}{q!}  \sum_{r = 0}^q \binom{q}{r}\sum_{p = 0}^{q-r} \binom{q-r}{p} K_1^p K_2^{q-r-p}     \\
    & \times 
    \sum_{l = 0}^r \binom{r}{l} 
    \left( {{\frac{K\Delta}{2}}} \right)^{2l}
  \frac{\Gamma (m_1+l+p)}{\left({m_1} \left(1+K\right)-(m_1+{K_1} ) \bar{\gamma }s\right)^{m_1+l+p}} \frac{\Gamma (m_2+l-p+q-r)}{\left({m_2} \left(1+K\right)-(m_2+{K_2} ) \bar{\gamma} s \right)^{m_2+l-p+q-r}} \left(\bar{\gamma } s\right)^{2 l-r} 
  \nonumber
    \\
    &\times
    {_2 \tilde{F}_1}\left(m_1+l+p, m_2+l-p+q-r;2l-r+1; \frac{(K\Delta\bar\gamma s)^2}{4(m_1(1+K)-(m_1+K_1)\bar\gamma s)(m_2(1+K)-(m_2+K_2)\bar\gamma s)} \right).
    \label{iftr_gmgf}\\ \nonumber \\ \hline \nonumber
    \end{align}
\end{figure*}

\begin{lemma}
	Let $\gamma \sim \mathcal{IFTR}(\overline{\gamma},m_1, m_2,K,\Delta)$, then its $n$-th order moment can be expressed as follows:
	
	(i) General case ($n\in \mathbb{R^+}$):
\begin{align}
	\mu_{\gamma}^n = \sum_{j=0}^{\infty} A_j \frac{ \Gamma(n+j+1) \bar\gamma^{n}}{\Gamma(j+1) (1+K)^n}.
\end{align}

  (ii) Case $n\in \mathbb{N}$: A closed-form expression is given now by
\begin{align} \label{moment}
  \mu_{\gamma}^n  &= \left( {\frac{{\overline \gamma  }}
{{1 + K}}} \right)^n \sum\limits_{q = 0}^n { \binom{n}{q} \frac{{n!}}
{{q!}}\sum\limits_{r = 0}^q  \binom{q}{r}  }  \hfill \nonumber\\&
 \times \sum\limits_{p = 0}^{q - r} {\binom{q-r}{p} K_1^p K_2^{q - r - p} \sum\limits_{l = 0}^r  }\binom{r}{l}  \left( {\frac{{K\Delta }}
{2}} \right)^{2l}  \hfill \nonumber\\&
  \times \frac{{\Gamma \left( {m_1  + l + p} \right)}}
{{\Gamma \left( {m_1 } \right)m_1^{l + p} }}\frac{{\Gamma \left( {m_2  + q - l - p} \right)}}
{{\Gamma \left( {m_2 } \right)m_2^{q - l - p} }}\delta _{2l,r}.
\end{align} 
\end{lemma}
where $\delta_{2l,r}$ is the kronecker delta function.
		
		\begin{IEEEproof}
These results follows by considering $s=0$ in the GMGF expressions. In case (ii), the following equality has been taken into account to obtain \eqref{moment}:
\begin{align} \label{s2f1}
\lim_{s\rightarrow 0}s^{n - m}  \cdot _2 \tilde{F}_1 \left( {a,b;n - m + 1;A \cdot s^2 } \right) = \delta _{n,m}, 
\end{align} 
which holds for any $n,m \in \mathbb{N}$, where the cases $n>m$ and $n=m$ are trivial, and the case $n<m$ results from the fact that the Gamma function has simple poles at the non-positive integers, and therefore from \eqref{regularized 2F1} and given $p \in \mathbb{N} \cup \{0\}$ we can write
\begin{align} 
_2 \tilde F _1 \left( {a,b; - p;z} \right) = \sum\limits_{k = p + 1}^\infty  {\frac{{\left( a \right)_k \left( b \right)_k }}
{{\Gamma \left( { - p + k} \right)}}} \frac{{z^k }}{{k!}}.
\end{align} 
\end{IEEEproof}

From the expression of the moments for $n\in \mathbb{N}$ given in \eqref{moment}, a closed-form expression for the amount of fading (AoF) for IFTR fading can be obtained in closed-form. The AoF captures the  severity, in terms of the variability, of the fading channel as a function of the parameters of the model and is defined as the variance of the SNR normalized by its squared mean, so that $\text{AoF} \triangleq E\{(\gamma -\bar\gamma)^2\}/\bar \gamma ^2=E\{\gamma^2\}/\bar \gamma ^2 - 1$.

\begin{corollary}
Let $\gamma \sim \mathcal{IFTR}(\overline{\gamma},m_1, m_2,K,\Delta)$, then, its AoF can be written as
\begin{align} \label{AoF}
\text{AoF} = \frac{1}
{{\left( {1 + K} \right)^2 }}\left[ {1 + 2K + \frac{{\left( {K\Delta } \right)^2 }}
{2} + \frac{{K_1^2 }}
{{m_1 }} + \frac{{K_2^2 }}
{{m_2 }}} \right].
\end{align} 

\end{corollary}
		\begin{IEEEproof}
This result is obtained by particularizing the moments in \eqref{moment} to the definition of the AoF.	
	\end{IEEEproof}
	
	The IFTR fading model tends to the TWDP one for $m_1,m_2 \rightarrow \infty$. As a check, it must be noted that for such condition the  expression given in \eqref{AoF} tends to the AoF given in \cite[eq. (34)]{rao2015mgf} for TWDP fading.

\section{Performance analysis}
By using the derived statistical characterization of the IFTR fading model, the performance of different wireless communication systems undergoing this fading distribution can be calculated. In the following, the channel capacity, the outage probability in an interference-limited multi-antenna receiver and the symbol error rate have been obtained for \gls{IFTR} fading.

\subsection{Average channel capacity}
The average capacity per unit bandwidth for \gls{IFTR} fading is given by
\begin{align}
	C = \int_{0}^{\infty} \log_2(1+x) f_{\gamma}^{\rm IFTR}(x) dx.
\end{align}
A direct application of Corollary I using the average channel capacity expression for Nakagami-$m$ fading channels \cite[eq. (23)]{alouini1997capacity} provides the following closed-form expression:
\begin{align}
	 C = \sum_{j=0}^{\infty} \frac{A_j e^{\frac{K+1}{\bar\gamma}}}{\ln(2)} \sum_{k=0}^{j} \left( \frac{1+K}{\bar\gamma}\right)^k \Gamma\left( -k,\frac{1+K}{\bar\gamma}\right),\label{C}
\end{align}
where $A_j$ is given in eq. \eqref{eq3} and $\Gamma(.,.)$ is the upper incomplete gamma function, which can
be computed, when the first parameter is a negative integer, as \cite[eq. (8.352.3)]{gradshteyn2014table}
\begin{align}
	\Gamma(-n,x)=\frac{(-1)^n}{\Gamma(n)}\left[ \sum_{r=0}^{n-1} \frac{\Gamma(n-r)}{(-x)^{n-r} e^{x}} - Ei(-x) \right],
\end{align}
where $E_i(\cdot)$ is the exponential integral function \cite[eq. (8.211.1)]{gradshteyn2014table}.

\subsection{Outage probability in interference-limited multi-antenna receiver}

The outage probability, i.e., the probability that the received SNR is below a threshold $\gamma_{th}$, under IFTR fading is given by
\begin{align}\label{pout}
P_{out} = Pr(\gamma<\gamma_{th}) = F_{\gamma}^{\rm IFTR}(\gamma_{th}).
\end{align}

On the other hand, in the presence of co-channel interference (CCI) of total received power $I$, considering negligible background noise and denoting as $W$ the received power from the desired user, which is assumed to experience IFTR fading, the outage probability is defined as
\begin{align}
	\hat P_{\rm out}= P\left(\frac{W}{I}<R_{th}\right),\label{pcci}
\end{align}
where $R_{th}$ denotes the signal-to-interference (SIR) threshold. 

We further assume $N$ receive antennas performing maximal ratio combining (MRC) and $L$ independent and identically distributed (i.i.d.) Rayleigh interferers with average power $P_I$. In this scenario, the outage probability is given by  \cite[eq. (15)]{CCI}
\begin{align}
	\hat P_{\rm out} =
	\sum_{k=0}^{L-1}\left(\frac{1}{ R_{th}P_I}\right)^k
	\sum_{\mathcal{U}} \prod_{i=1}^N \frac{1}{u_i !} {\phi^{(u_i)}_{W_i}\left(-\frac{1}{ R_{th}P_I}\right)},
 \label{hPout}
\end{align}
where $\mathcal{U}$ is a set of $N$-tuples such that $\mathcal{U}=\{(u_1 ... u_N), u_i\in \mathbb{N},  ~\sum_{i=1}^N u_i = k\}$, 
and $\phi^{(u_i)}_{W_i}(s)$ is computed using \eqref{iftr_gmgf}, as $u_i \in \mathbb{N}$, by simply considering the relation $W_i = \frac{\gamma_i}{E_s/\overline N_0}$, thereby providing a closed-form expression for the outage probability.

\subsection{Exact and approximated average BER}

The average symbol error rate in a telecommunication system is one of the main parameters for measuring the quality of communication. In this section, we calculate this metric for the \gls{IFTR} fading channel. The conditional BER probability in AWGN channel for some relevant modulations with coherent detection can be written as \cite{5415562}
\begin{align}\label{eq15}
	P_e(x) = \sum_{r=1}^{R} \alpha_r Q(\sqrt{\beta_r x}).
\end{align}
The average BER is calculated by averaging over all possible channel realizations. From the result in \cite[eq. (5.18)]{Simon05} for Nakagami-$m$ fading, by virtue of Corollary 1, the average BER in IFTR fading can be written, after some manipulation, as
\begin{align}
	\bar{P}_e = &
  \sum\limits_{r = 1}^R {\frac{{\alpha _r }}{2}} \sum\limits_{j = 0}^\infty  {A_j } \left[ {1 - \sqrt {\frac{{\beta _r \overline \gamma  }}
{{2\left( {1 + K} \right) + \beta _r \overline \gamma  }}}  \sum\limits_{k = 0}^j \binom{2k}{k} } \right. \hfill \nonumber\\&
  \left. { \times \left( {\frac{{1 - \frac{{\beta _r \overline \gamma  }}
{{2\left( {1 + K} \right) + \beta _r \overline \gamma  }}}}
{4}} \right)^k } \right]. \hfill 
\end{align}

In the high SNR regime ($\bar{\gamma} \to \infty$), the average BER can be simplified by simply maintaining the first term in the infinite summation, as stated in Remark 2, yielding
\begin{align}
  \bar{P}_e \approx \sum\limits_{r = 1}^{R} {\frac{{\alpha _r }}
{2}} A_0 \left[ 1 - \sqrt {\frac{{\beta _r \overline \gamma  }}
{{2\left( {1 + K} \right) + \beta _r \overline \gamma  }}} \right], \ \bar{\gamma} \to \infty. \hfill 
\end{align}

\section{Numerical and simulation results}
This section presents figures illustrating the performance of IFTR fading channels. The obtained numerical results have been validated by Monte Carlo simulations where $10^7$ random realizations of the IFTR distribution have been computed.
Based on Table I, numerical results involving infinite series have been calculated truncating to 40 terms, as it provides a satisfactory accuracy for all the considered cases. In all the presented figures we have assumed $\bar\gamma = 1$.

\begin{figure}[t]
    \centering
    \includegraphics[height=7cm]{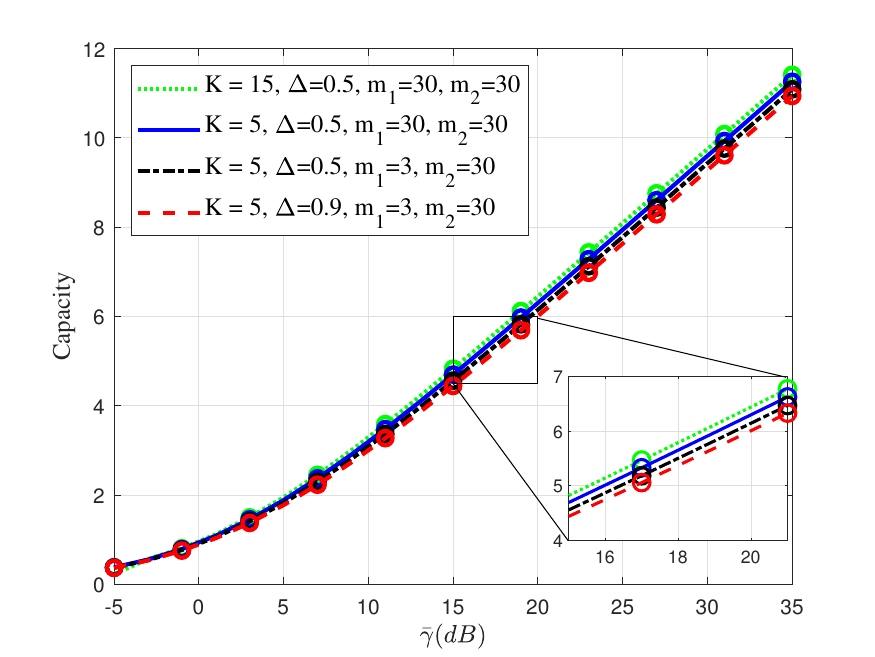}
    \caption{Numerical and simulation results for the average capacity vs. average SNR in dB for different channel parameters values and ($\bar\gamma=1$). Simulation confirmation results are displayed as circular markers.}
    \label{figure4}
\end{figure}

\begin{figure}[t]
    \centering\includegraphics[height=7cm]{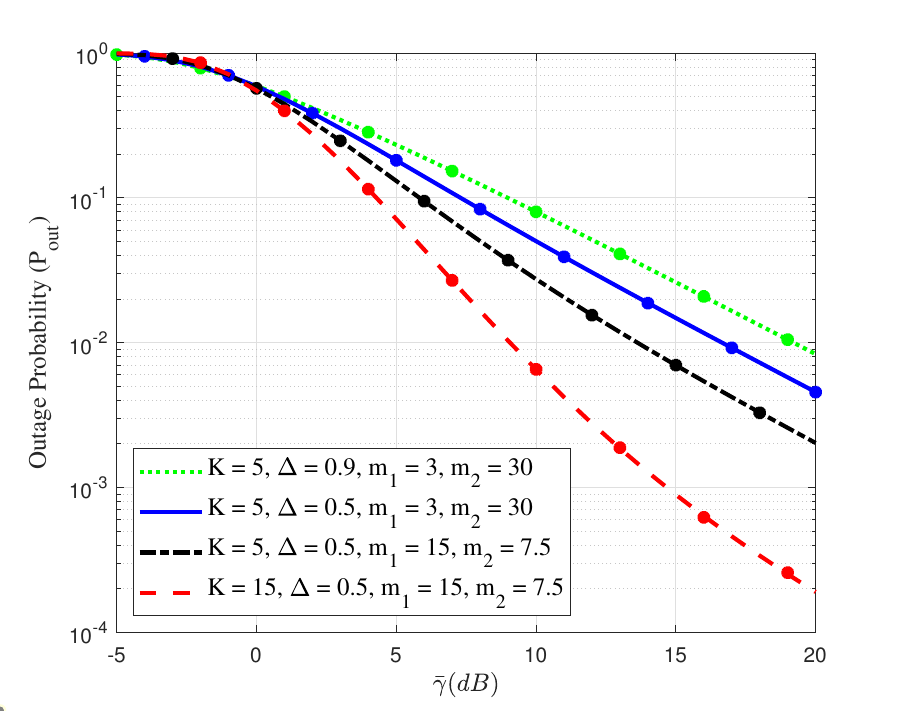}
    \caption{Numerical and simulation results for the outage probability vs. $\bar{\gamma}$ with $\gamma_{th} = 0$ dB. Simulation confirmation results are displayed as circular markers.}
    \label{figure5}
\end{figure}

\begin{figure}[t]
    \vspace{-0.5mm}
    \centering
    \includegraphics[height=7cm]{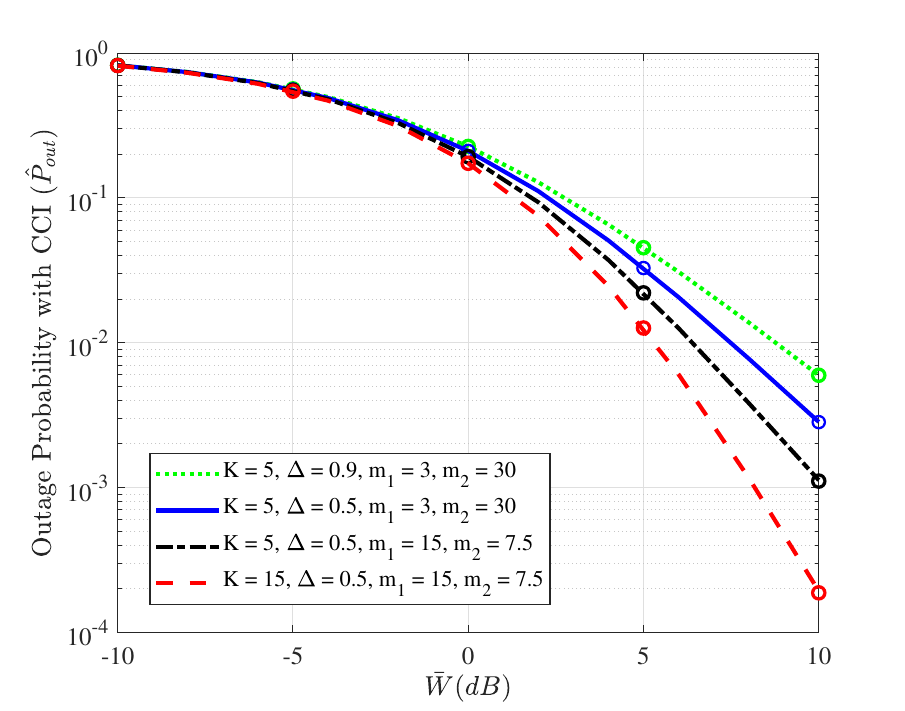}
    \caption{Numerical and simulation results for the outage probability for IFTR fading with CCI for different channel parameters values with $N=2$, $L=1$, $P_I=1$ and $R_{th}=0$ dB. Simulation confirmation results are displayed as circular markers. }
    \label{figure6}
\end{figure}

The average capacity for \gls{IFTR} fading is presented in Fig. \ref{figure4} for different values of the channel parameters  $\{m_1, m_2,K,\Delta\}$. The presented numerical results have been obtained from \eqref{C}. It can be seen that a higher capacity is obtained for high $K$. On the other hand, a high value of $\Delta$ (close to 1) yields lower capacity due to the increased probability that the specular components cancel each other, which increases the channel variability. 

Fig. \ref{figure5} shows the outage probability ($P_{out}$) computed from \eqref{pout} versus the average SNR ($\bar\gamma$) for different channel model parameters. It can be observed that decreasing $\Delta$ from $0.9$ to $0.5$, increasing $K$ from 5 to 15, and decreasing $m_1$ and $m_2$ yields a better performance (lower outage probability), as these changes give rise to a reduced fading severity. 

In Fig. \ref{figure6} the same values of the IFTR model parameters as in Fig. \ref{figure5} are considered, although now a multiantenna receiver is assumed under CCI for the outage probability. The same effect as in  Fig. \ref{figure5} is observed when the channel parameters are modified, but the amount of variation in the outage probability is affected by the presence of CCI and the use of MRC reception. 
For example, for $\bar \gamma =\overline {SIR}=10$ dB, the outage probability under CCI, $\hat P_{out}=2\times 10^{-4}$, is lower than $P_{out}=5 \times 10^{-3}$ due to the MRC diversity gain when the values of the parameters are $K=15$, $\Delta=0.5$, $m_1=15$ and $m_2=7.5$.

\begin{figure}[t]
	\centering
	\includegraphics[width=1.01\linewidth]{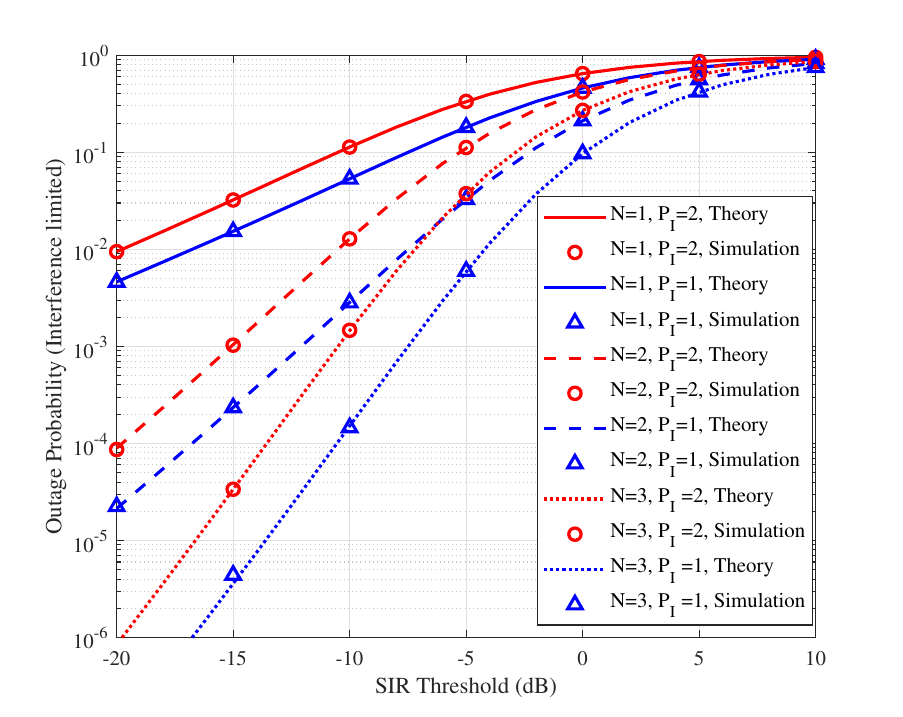}
	\caption{Numerical and simulation results for the outage probability for IFTR fading with CCI considering different values of $N=1,2,3$ and  interference average power $P_I=1,2$ watts. Channel parameters are $m_1=3$, $m_2 =30$, $K=5$ and $\Delta =0.5$. }
	\label{figure7}
\end{figure}

Fig. \ref{figure7} shows the outage probability with CCI for different system parameters vs. the SIR threshold. The numerical results of the outage probability from \eqref{hPout} are plotted for different number of antennas $N=1,2,3$ and average interference power $P_I=1,2$ when $L=1$. It can be seen that as the number of received antennas increases, the outage probability decreases, and the diversity gain increases. Also, for a given SIR threshold, the outage probability is higher for larger average interference power, as expected. Monte-Carlo simulations show an excellent match to the numerical results.

\begin{figure}[t]
    \centering
    \includegraphics[height=7cm]{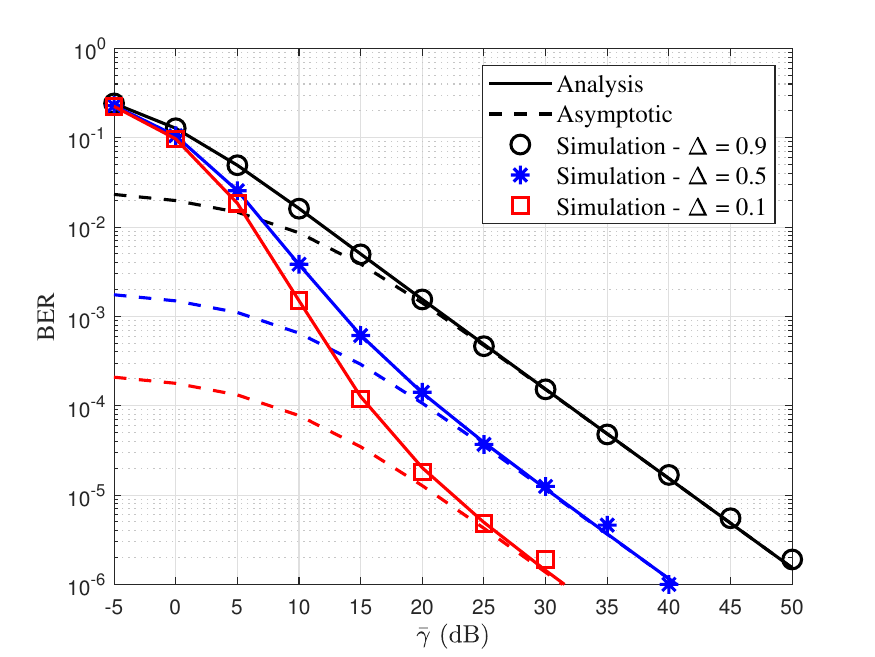}
    \caption{Numerical and simulation results for the average BER vs. average SNR in dB for BPSK with channel parameters $m_1=15.7$, $m_2=5.1$ and $K=10$. }
    \label{figure8}
\end{figure}

Finally, Fig. \ref{figure8}  shows the exact and asymptotic BER vs. the average SNR in IFTR fading for BPSK modulation ($R=1$, $\alpha_1=1$, $\beta_1=2$). The figure shows this performance metric for different channel parameters $\Delta=0.1,0.5,0.9$ when the fluctuating parameters $m_1$($=15.7$) and $m_2$($=5.1$) are non-integers. Again, increasing $\Delta$ results in higher channel variability, causing a detrimental impact on performance, i.e., a higher average BER. It is worth mentioning that when the average SNR is above 20 dB, the asymptotic curves, which are much simpler to compute, yield very good approximated results, and above 30 dB the exact and asymptotic results are indistinguishable in all the presented cases.

%

\section{Conclusion} 
In this paper, a new formulation in series form has been derived for the  PDF and CDF of the \gls{IFTR} fading model. The convergence of the obtained series are demonstrated and truncated for numerical computation using the Kolmogorov-Smirnov goodness-of-fit.
We show that, leveraging on any average performance metric already known for the much simpler Nakagami-$m$ fading model, such metric can be readily obtained for IFTR fading. 

Additionally, the GMGF of IFTR fading has been obtained which, for most cases of interest, can be expressed in closed-form, thus opening the door to circumvent the model mathematical complexity and obtain several relevant performance metrics also in closed-form, as well as the moments of the distribution and the amount of fading.
Finally, the new and expanded statistical characterization of the IFTR fading model has been exemplified, showing and discussing numerical results for  the average capacity, the outage probability with and without interference and the BER for BPSK modulation, which have been verified by Monte Carlo simulations.

\section*{Acknowledgment}
The authors would like to express their gratitude to F. Javier López-Martínez for his insightful comments  during the development of this work.

\appendices
\section{Proof of Lemma I} \label{A1}
Let us consider the fading model defined in (\ref{eq1}) conditioned to the particular realizations of the RVs $\zeta_1=u_1$, $\zeta_2=u_2$. Thus, we can write
\begin{equation}
\label{eq:12}
\left. {V_r } \right|_{ u_1, u_2}  = \sqrt u_1 V_1 e^ {j\phi _1 }  + \sqrt u_2 V_2 e^ {j\phi _2 }  + X + jY,
\end{equation}
which corresponds to the TWDP fading model with specular components amplitudes $\sqrt u_1 V_1$ and $\sqrt u_2 V_2$ and  parameters
\begin{equation}
\label{eq:13}
K_{u_1 ,u_2 }  = \frac {u_1 V_1^2  + u_2 V_2^2 } {2\sigma ^2}=u_1K_1+u_2K_2,
\end{equation}
\begin{equation}
\label{eq:14}
\Delta _{u_1 ,u_2 }  =  \frac {{2\sqrt {u_1 u_2 } V_1 V_2 }}  {u_1 V_1^2  + u_2 V_2^2 },
\end{equation}
which satisfy
\begin{equation}
\label{eq:23a}
\begin{split}
  & K_{u_1 ,u_2 }\Delta _{u_1 ,u_2 }   = \sqrt {u_1 u_2 } \frac { V_1 V_2 }  {\sigma ^2 }= \sqrt {u_1 u_2 }K\Delta.
\end{split}
\end{equation}
The conditional average SNR for the model definition given in (\ref{eq:12}) will be
\begin{equation}
\label{eq:17}
\begin{split}
  & \bar \gamma _{u_1 ,u_2 }  = \frac{E_s}{N_0} \left( {u_1 V_1^2  + u_2 V_2^2 + 2\sigma ^2 } \right) = \frac{E_s}{N_0} 2\sigma ^2 \left( {1 + K_{u_1 ,u_2 } } \right). 
\end{split}
\end{equation}
On the other hand, by promediating over all possible realizations of the unit-mean RVs $\zeta_1$, $\zeta_2$, the unconditional average SNR will be 
\begin{equation}
\label{eq:17b}
\bar\gamma = E\{\bar\gamma _{u_1 ,u_2 }\}=\frac{E_s}{N_0}(V_1^2+V_2^2+2\sigma^2)= \frac{E_s}{N_0}2\sigma^2(1+K),
\end{equation}
and therefore, equating \eqref{eq:17} and \eqref{eq:17b}, we can write
\begin{equation}
\label{eq:19}
\frac {1 + K_{u_1 ,u_2 }} {\bar \gamma _{u_1 ,u_2 } } = \frac {1}{\left( {E_s /N_0 } \right)2\sigma ^2 } = \frac{1 + K}{\bar \gamma },
\end{equation}

From the PDF of the received power of the TWDP fading model given in \cite{ermolova2016capacity} as a mixture of Gamma distributions, the PDF of the conditional SNR of the model defined in (\ref{eq:12}) can be written as
\begin{align}\nonumber
	&f_ {\gamma_{u_1 ,u_2}}^{\text{TWDP}} (x)= e^{ - {K_{u_1 ,u_2}}}\sum\limits_{j = 0}^\infty  {\frac{{K_{u_1 ,u_2}^j}}{{j!}}} {f^\mathcal{G}}\left( x;j + 1,\frac {\bar \gamma _{u_1 ,u_2 } }{1 + K_{u_1 ,u_2 }}  \right)
	\\& \times\sum\limits_{k = 0}^j \binom{j}{k} {\left( {\frac{{{\Delta _{u_1 ,u_2}}}}{2}} \right)^k}\sum\limits_{l = 0}^k \binom{k}{l} {I_{2l - k}}\left( { - {K_{u_1 ,u_2}}{\Delta _{u_1 ,u_2}}} \right),
\end{align}
which, from \eqref{eq:13}-\eqref{eq:19}, can be rewritten as
\begin{align}\nonumber
	&f_ {\gamma_{u_1 ,u_2}}^{\text{TWDP}} (x) =	e^{ - u_1{K_1} - u_1{K_2}} \sum\limits_{j = 0}^\infty  {\frac{1}{{j!}}} {f^\mathcal{G}}\left( x;j + 1,\frac {\bar \gamma }{1 + K} \right)\\
	&\times \sum\limits_{k = 0}^j \binom{j}{k} \sum\limits_{q = 0}^{j - k} \binom{j-k}{q}{\left( {u_1{K_1}} \right)^q} {\left( {u_2{K_2}} \right)}^{j - k - q} \nonumber\\
	&  \times {\left( {\frac{{\sqrt {u_1 u_2} K\Delta }}{2}} \right)^k}\sum\limits_{l = 0}^k \binom{k}{l} {I_{ {2l - k}}}\left( -{\sqrt {u_1 u_2}K\Delta } \right),\label{eq:20}
\end{align}

The PDF of the SNR of the IFTR model can be obtained by averaging (\ref{eq:20}) over all possible realizations of the RVs $\zeta_1$ and $\zeta_2$, i.e. 
\begin{equation}
\label{eq:26}
\begin{split}
f_ {\gamma}^{\text{IFTR}} (x) = \int_0^\infty  {\int_0^\infty  f_ {\gamma_{u_1 ,u_2}}^{\text{TWDP}} (x) f_{\zeta _1 } \left( {u_1 } \right)f_{\zeta _2 } \left( {u_2 } \right)du_1 du_2 },
\end{split}
\end{equation}
where
\begin{equation}
\label{eq:02}
f_{\zeta _i} \left( u_i \right) = \frac{m_i ^{m_i } u_i^{m_i  - 1}}{\Gamma \left( {m_i } \right)} e^{ - m_i u_i} ,\quad \quad i = 1,2.
\end{equation}

The double integral in  (\ref{eq:26}) can be solved in closed-form by iteratively integrating with respect to variables $u_1$ and $u_2$. Thus, after changing the order of integration and summation, we can write
\begin{align}
\label{eq:27}
  f_\gamma ^{{\text{IFTR}}} (x) &= \sum\limits_{j = 0}^\infty  {f^\mathcal{G} } \left( {x;j + 1,\frac{{\overline \gamma  }}
{{1 + K}}} \right) \hfill \nonumber\\&
   \times \sum\limits_{k = 0}^j \binom{j}{k} \sum\limits_{q = 0}^{j - k}  \binom{j-k}{q}\frac{{K_1^q K_2^{j - k - q} }}
{{j!}} \hfill \nonumber\\&
   \times \left( {\frac{{K\Delta }}
{2}} \right)^k \sum\limits_{l = 0}^k { \binom{k}{l} \frac{{m_1^{m_1 } }}
{{\Gamma (m_1 )}}\frac{{m_2^{m_2 } }}
{{\Gamma (m_2 )}}} \mathcal{H}_1,
\end{align}
where we have defined 
\begin{align}
\label{eq:28}
  &\mathcal{H}_1 \triangleq \int_0^\infty  {u_2^{m_2  + j - k/2 - q - 1} } e^{ - \left( {m_2  + K_2 } \right)u_2 } \mathcal{I}_1(u_2 )du_2,   \\&
  \mathcal{I}_1(u_2 ) \triangleq \int_0^\infty  {u_1^{m_1  + q + k/2 - 1} } e^{ - \left( {m_1  + K_1 } \right)u_1 } \nonumber \\& \ \ \ \ \ \ \ \times I_{2l - k} \left( { - \sqrt {u_1 u_2 } K\Delta } \right)du_1. \label{eq:28b}
\end{align}
We now consider the following equality from \cite[6.643.2]{gradshteyn2014table} and \cite[9.220.2]{gradshteyn2014table}:
\begin{align}
	\mathcal{J} &= \int_0^\infty  {{t^{\mu  - 1/2}}{e^{ - pt}}{I_{2\nu }}\left( {2\beta  \sqrt t } \right)} dt 
	\nonumber \\ &
	=\frac{\Gamma(\mu+\nu+\frac{1}{2})\beta^{2\nu}}{p^{\nu+\mu+\frac{1}{2}}} \;
_1 \tilde{F}_1 \bigg(\mu+\nu+\frac{1}{2},2\nu+1,\frac{\beta^2}{p} \bigg),
\end{align}
where $_1 \tilde{F}_1$ is the regularized Kummer hypergeometric function, and from which \eqref{eq:28b} can be written in closed-form as
\begin{align}
\label{eq:29}
 \mathcal{I}_1(u_2 ) & = \left( { - 1} \right)^{k}\frac{{\Gamma \left( {m_1  + q + l} \right)}}
{{\,\left( {m_1  + K_1 } \right)^{m_1  + q + l} }} \left( {\frac{{K\Delta }}
{2}} \right)^{2l - k} u_2^{l - k/2}  \nonumber \\&
   \times _1\tilde F _1 \left( {m_1  + q + l;2l - k + 1;\frac{{u_2 K^2 \Delta ^2 }}
{{4\left( {m_1  + K_1 } \right)}}} \right).
\end{align}
Introducing \eqref{eq:29} into \eqref{eq:28} and  solving the integral with the help \cite[eq. (7.621.4)]{gradshteyn2014table} we can write 
\begin{align}
\label{eq:30}
  &\mathcal{H}_1 
   = \left( { - 1} \right)^k \left( {\frac{{K\Delta }}
{2}} \right)^{2l - k} \frac{{\Gamma \left( {m_1  + q + l} \right)}}
{{\,\left( {m_1  + K_1 } \right)^{m_1  + q + l} }} \hfill \nonumber \\&
   \times \frac{{\Gamma (m_2  + j - k - q + l)}}
{{\left( {m_2  + K_2 } \right)^{m_2  + j - k - q + l} }} \,_2 \tilde F_1 \bigg( {m_1  + q + l,} \bigg. \hfill \nonumber \\&
  \left. {m_2  + j - k - q + l;2l - k + 1;\frac{{K^2 \Delta ^2 }}
{{4\left( {m_1  + K_1 } \right)\left( {m_2  + K_2 } \right)}}} \right),
\end{align}
which, together with \eqref{eq:27}, yields the desired result in \eqref{pdfI} for the PDF of the SNR of the IFTR fading model. On the other hand, the CDF in \eqref{cdfI} is obtained by a simple integration of \eqref{pdfI} (see additional comments on this in Section \ref{KS}).


\section{Proof of Lemma 2: Case (ii)}\label{app1}
As in Appendix A, we consider an IFTR model conditioned to the particular realizations of the RVs $\zeta_1=u_1$, $\zeta_2=u_2$, which yields a TWDP model with specular components amplitudes $\sqrt u_1 V_1$ and $\sqrt u_2 V_2$, parameters $K_{u_1 ,u_2 }$ and $\Delta _{u_1 ,u_2}$ given, respectively, by \eqref{eq:13} and \eqref{eq:14}, and conditional mean $\bar \gamma _{u_1 ,u_2}$, given in \eqref{eq:17}.
The \gls{GMGF} for the TWDP model for $n \in \mathbb{N}$ can be obtained from \cite[eq. (24) for $\mu=1$]{MTWconference} as
\begin{align}\label{eq:023}
    &	\phi _{\bar\gamma_{u_1 ,u_2}} ^{(n)} \left( s \right) = \bar \gamma_{u_1 ,u_2}  ^n n! e^{\frac{{ K_{u_1 ,u_2}\bar \gamma_{u_1 ,u_2}  s}}{{ {1 + K_{u_1 ,u_2}} - \bar \gamma_{u_1 ,u_2}  s}}}  \sum\limits_{q = 0}^n \binom{n}{q} \frac{{K_{u_1 ,u_2}^q}}{q!}\nonumber \\
    & \times \frac{{\left( 1 + K_{u_1 ,u_2} \right)^{q+1} }}{{\left( {1 + K_{u_1 ,u_2} - \bar \gamma_{u_1 ,u_2}  s} \right)^{q+n+1} }} \hfill \sum\limits_{r = 0}^q \binom{q}{r} \left( {\frac{\Delta_{u_1 ,u_2} }{2}}\right)^r \nonumber \\
    & \times \sum\limits_{l = 0}^r \binom{r}{l} I_{2l - r} \left( {\frac{{ K_{u_1 ,u_2}\Delta_{u_1 ,u_2} \bar \gamma_{u_1 ,u_2}  s}}{{1 + K_{u_1 ,u_2} - \bar \gamma_{u_1 ,u_2}  s}}} \right),
\end{align}
which can be written, by using the relations \eqref{eq:13}-\eqref{eq:19}, as
\begin{align}\label{eq:123}
  &\phi _{\gamma _{u_1 .u_2 } }^{{\text{(n)}}} (s) = \bar \gamma  ^n n!e^{\frac{{\overline \gamma  s}}
{{1 + K - \bar \gamma  s}}\left( {u_1 K_1  + u_2 K_2 } \right)} \sum\limits_{q = 0}^n \binom{n}{q} \sum\limits_{p = 0}^{q - r} \binom{q-r}{p}   \nonumber \\&\times
  \frac{{\left( {u_1 K_1 } \right)^p \left( {u_2 K_2 } \right)^{q - r - p} }}
{{q!}}\frac{{\left( {1 + K} \right)^{q + 1} }}
{{\left( {1 + K - \bar \gamma  s} \right)^{q + n + 1} }}\sum\limits_{r = 0}^q \binom{q}{r}  \hfill \nonumber \\&\times
  \left( {\frac{{\sqrt {u_1 u_2 } K\Delta }}
{2}} \right)^r \sum\limits_{l = 0}^l \binom{r}{l} I_{2l - r} \left( {\sqrt {u_1 u_2 } \frac{{K\Delta \bar \gamma  s}}
{{1 + K - \bar \gamma  s}}} \right).
\end{align}

\begin{figure*}
\begin{align} 
\label{eq:100}
  \mathcal{H}_2 =& \left( {1 + K - \overline \gamma  s} \right)^{m_1  + m_2  + q} \left( {\overline \gamma  s} \right)^{2l - r} \left( {\frac{{K\Delta }}
{2}} \right)^{2l - r}  \nonumber  \frac{{\Gamma \left( {m_1  + l + p} \right)}}
{{\,\left( {m_1 \left( {1 + K - \overline \gamma  s} \right) - K_1 \overline \gamma  s} \right)^{m_1  + l + p} }}\nonumber \\&
   \times
\frac{{\Gamma (m_2  + l - p + q - r)}}
{{\left( {m_2 \left( {1 + K - \overline \gamma  s} \right) - K_2 \overline \gamma  s} \right)^{m_2  + l - p + q - r} }}  {_2 \tilde{F}_1} \bigg( m_1  + l + p,m_2  + l - p + q - r;2l - r; \bigg. \nonumber \\&
   	\left. \frac{{\left( {K\Delta \overline \gamma  s} \right)^2 }}
{{4\left( {m_2 \left( {1 + K - \overline \gamma  s} \right) - K_2 \overline \gamma  s} \right)\left( {m_1 \left( {1 + K - \overline \gamma  s} \right) - K_1 \overline \gamma  s} \right)}} \right). \tag{80} \\ \nonumber \\ \hline \nonumber
\end{align}
\end{figure*}

The GMGF of IFTR fading is obtained by averaging \eqref{eq:123} over all possible realizations of $\zeta_1, \zeta_2$ as
\begin{align}
    \phi _{\gamma} ^{(n)} \left( s \right) =\int_{0}^{\infty} & \int_{0}^{\infty} \phi _{\gamma _{u_1 .u_2 } } ^{(n)}\left(s\right)  f_{\zeta_1}(u_1)  f_{\zeta_2}(u_2)du_1 du_2.\label{g22}
\end{align}
Introducing \eqref{eq:123} into \eqref{g22} we can write 
\begin{align}\label{eq:123b}
  &\phi _{\gamma _{u_1 .u_2 } }^{{\text{(n)}}} (s) = \bar \gamma  ^n n!\frac{{m_1^{m_1 } }}
{{\Gamma (m_1 )}}\frac{{m_2^{m_2 } }}
{{\Gamma (m_2 )}}\sum\limits_{q = 0}^n \binom{n}{q} \sum\limits_{p = 0}^{q - r} \binom{q-r}{p}  \hfill \nonumber \\&\times
  \frac{{\left( {K_1 } \right)^p \left( {K_2 } \right)^{q - r - p} }}
{{q!}}\frac{{\left( {1 + K} \right)^{q + 1} }}
{{\left( {1 + K - \bar \gamma  s} \right)^{q + n + 1} }}\sum\limits_{r = 0}^q \binom{q}{r}  \hfill \nonumber \\&\times
  \left( {\frac{{K\Delta }}
{2}} \right)^r \sum\limits_{l = 0}^l  \binom{r}{l} \mathcal{H}_2,   
\end{align}
where we have defined
\begin{align}
\label{eq:28_2}
  &\mathcal{H}_2 \triangleq \int_0^\infty  {u_2^{m_2  + q - r/2 - p - 1} } e^{ - \left( {m_2  - \frac{{K_2 \bar \gamma  s}}
{{1 + K - \bar \gamma  s}}} \right)u_2 } \mathcal{I}_2 (u_2 )du_2, \\& 
  \mathcal{I}_2 (u_2 ) \triangleq \int_0^\infty  {e^{ - \left( {m_1  - \frac{{K_1 \bar \gamma  s}}
{{1 + K - \bar \gamma  s}}} \right)u_1 } u_1^{m_1  + p + r/2 - 1} }  \hfill \nonumber\\&
  \quad \quad \quad \quad \quad \quad I_{2l - r} \left( {\sqrt {u_1 u_2 } \frac{{K\Delta \bar \gamma  s}}
{{1 + K - \bar \gamma  s}}} \right)du_1. \label{eq:28b_2}
\end{align}
Note that $\mathcal{H}_2$ and $\mathcal{I}_2$ are actually the same integrals $\mathcal{H}_1$ and $\mathcal{I}_1$ defined, respectively, in \eqref{eq:28} and \eqref{eq:28b}, although for different coefficients, which are now in some cases rational functions on $s$. Therefore, following the same procedure as in \eqref{eq:28}-\eqref{eq:30}, a closed-form expression can be found for $\mathcal{H}_2$ as given in \eqref{eq:100}, which together with \eqref{eq:123b} yields \eqref{iftr_gmgf}.

\bibliographystyle{IEEEtran}
\bibliography{ReferencesBIB}

\end{document}